\documentclass[runningheads,a4paper]{llncs}
\usepackage{amssymb}
\usepackage{amsmath}
\usepackage{graphicx}
\usepackage{url}
\newcommand{\keywords}[1]{\par\addvspace\baselineskip
\noindent\keywordname\enspace\ignorespaces#1}
\begin{document}

\mainmatter

\title{Machine Learning for Mathematical Software}  
\titlerunning{Machine Learning for Mathematical Software} 
\author{Matthew England}
\authorrunning{Matthew England}
\institute{
Coventry University, UK\\
\email{Matthew.England@coventry.ac.uk}\\ 
}
\maketitle

\begin{abstract}
While there has been some discussion on how Symbolic Computation could be used for AI there is little literature on applications in the other direction. However, recent results for quantifier elimination suggest that, given enough example problems, there is scope for machine learning tools like Support Vector Machines to improve the performance of Computer Algebra Systems. We survey the authors own work and similar applications for other mathematical software.

It may seem that the inherently probabilistic nature of machine learning tools would invalidate the exact results prized by mathematical software. However, algorithms and implementations often come with a range of choices which have no effect on the mathematical correctness of the end result but a great effect on the resources required to find it, and thus here, machine learning can have a significant impact.
\keywords{Machine Learning, Mathematical Software}
\end{abstract}

\section{Introduction}

Machine Learning, refers to tools that use statistical techniques to give computer systems the ability to \emph{learn} rules from data; that is, improve their performance on a specific task, without changing their explicit programming.  Although many of the core approaches date back decades, machine learning has found great success in recent years, driven by the advances in both computer hardware and the availability of data. There have well publicised successes of machine learning recently such as Google's \textsc{AlphaGo} being the first to beat a professional human Go player\footnote{\scriptsize \url{https://research.googleblog.com/2016/01/alphago-mastering-ancient-game-of-go.html}}.  We are all likely to have interactions with software that at least partially learns on a daily basis, whether through traffic signal control \cite{YQKLK17} or the extraction and interpretation of our views \cite{YSZ17}.

Most industries have felt some effect from the advance of these tools, and software engineering itself is no different.  Indeed, the idea of using machine learning in the software development process is not a new one \cite{ZT03}.  In particular, machine learning is now a common tool in the testing and security analysis of software \cite{GS17}.  
Machine learning is at its most attractive when the underlying functional relationship to be modelled is complex or not well understood.  It may seem that machine learning is hence not relevant to the sub-field of mathematical software where underlying functional relationships are the key object of study.  Further, the inherently probabilistic nature of machine learning tools seems like it would invalidate the exact mathematical results prized by such software.  

However, as most developers would acknowledge, mathematical software often comes with a range of choices which, while having no effect on the correctness of the end result, could have a great effect on the resources required to find it.  These choices range from the low level (in what order to perform a search that may terminate early) to the high (which of a set of competing exact algorithms to use for this problem instance).  In making such choices we may be faced with decisions where the underlying relationships are not fully understood, but are not themselves the key object of study.  Thus in practice we will use a, usually fairly crude, man-made heuristic in order to proceed with the implementation.  

It is possible that many of these decisions could be improved by allowing learning algorithms to analyse the data.  It is even possible that such study could lead to a better understanding of the underlying relationship.  For example, a standard step in the use of machine learning is feature selection: identifying a minimal number of features about the data to use in making the decision.  The primary reason for this is to reduce the resources required to train a classifier, and reduce the risk of over-fitting.  However, in identifying the most important features the developers of mathematical software may also get insight on new mathematical results, or at least hypotheses to guide future development.

We proceed by surveying the author's own work applying machine learning in one particular area of symbolic computation.  We then consider where else in computer algebra and mathematical software more broadly there may be potential applications and existing inspiration. 

\section{Machine Learning for CAD}
\label{SEC:MyWork}

The author has been involved in two applications of machine learning \cite{HEDP16}, \cite{HEWDPB14} to improve the performance of a particular algorithm.  \emph{Cylindrical Algebraic Decomposition} (CAD) refers to both a mathematical object and the algorithms to produce them, both first introduced by Collins in the 1970s.  Here:
\begin{itemize}
\item \emph{decomposition} means a partition of $\mathbb{R}^n$ into connected subsets called \emph{cells};
\item \emph{algebraic} is short for \emph{semi-algebraic} and means that each cell may be described by a conjunction of polynomial constraints;
\item \emph{cylindrical} refers to the structure of the decomposition: the projections of any two cells, onto a lower coordinate space with respect to the given variable ordering, are either identical or disjoint.  
\end{itemize}
CADs were originally produced as \emph{sign-invariant} for a set of input polynomials\footnote{See for example \cite{ACM84I} for a description of the original CAD algorithm.}, meaning each polynomial is to have constant sign on each cell.  However, for almost all applications what is truly required is a decomposition \emph{truth-invariant} for logical formulae: where each formula has constant truth value on each cell.  A sign-invariant decomposition for the polynomials in the formulae produces truth invariance, but it can be achieved more efficiently \cite{McCallum1999b}, \cite{EBD15}, \cite{BDEMW16}.  

In either case, the invariance properties mean only a finite number of sample points need to be queried to solve problems.  In particular, CADs offer a tool to perform Quantifier Elimination (QE). Through QE there are a multitude of applications throughout engineering and the sciences (see for example \cite{Sturm2017}).  Additional application of CAD directly include identification of steady states in biological networks \cite{BDEEGGHKRSW17}, and programming with multi-valued functions \cite{DBEW12}.

However, CAD is well known for its worst case complexity doubly exponential \cite{DH88}\footnote{Doubly exponential usually in the number of variables, although the logical structure can be used to improve this somewhat \cite{BDEMW16}, \cite{EBD15}, \cite{ED16a}.}.  Hence it is important to optimise how CAD is used, such as the setting of any optional parameters and the presentation of input.

\subsection{Deciding whether to pre-condition}
\label{SUBSEC:GB}

One choice a user could make is whether to give their problem to CAD directly, or to first precondition it.  One common technique for input formulae with multiple equations is the use of a \emph{Gr\"obner Basis} (GB).  A GB is a particular generating set of an ideal with useful properties: although our task is not to study the ideal it turns out the GB can give a simpler representation for CAD to work with.

To be precise: let $E = \{e_1, e_2, \dots\}$ be a set of polynomials; 
$G = \{g_1, g_2, \dots\}$ be a GB for $E$; and $B$ be any Boolean combination of constraints. 
Then 
\begin{align*}
\Phi &= (e_1 = 0 \land e_2 = 0 \land \dots) \land B \mbox{ and } \\
\Psi &= (g_1 = 0 \land g_2 = 0 \land \dots) \land B
\end{align*}
are equivalent, and a CAD truth-invariant for $\Psi$ can solve problems involving  $\Phi$.

This was studied first in 1991 \cite{BH91} and then again in 2012 \cite{WBD12_GB}.  In both cases the conclusion was that usually GB pre-conditioning is beneficial for CAD, but there are some examples where it is greatly detrimental.  
In \cite{HEDP16} we considered using machine learning to decide when to use GB.  On a dataset of over a thousand randomly generated problems with multiple equations we found 75\% were easier to study after a GB was taken.  We trained a \emph{Support Vector Machine} (SVM) classifier \cite{CV95} with radial basis function (see for example \cite{SC04}) to make the decision.  We used as problem features simple algebraic properties (degrees, density of occurrence of variables etc.) of both the input polynomials and the GB.  Only when including those of the GB could the classifier make good decisions: not a problem since for any problem where CAD is tractable GB is trivial.  The classifier chooses, not whether to construct the basis, but whether to use it.  In \cite{HEDP16} we also showed how feature selection experiments (identifying a minimal subset of the features) could improve accuracy (reducing the risk of over-fitting).

\subsection{Choosing a variable ordering}
\label{SUBSEC:VarOrd}

Another choice a user may have to make for CAD is the variable ordering, used in the definition of cylindricity, and crucial to the computational path of the algorithm.  Depending on the application this may be free, constrained or fixed.  

For example, for QE one must order variables as they are quantified; but there is no restriction on free variables and adjacent quantifiers of the same type may be swapped.
It is well known that this choice can dramatically affect the feasibility of a problem.  In fact, there are a class of problems in which one variable ordering gives output of double exponential complexity in the number of variables and another output of a constant size \cite{BD07}.  There are heuristics available to make the choice but each can be misled by certain examples.

In \cite{HEWDPB14} we investigated machine learning for this choice.  In this case the choice is not binary but from many different orderings\footnote{If the choice is completely free then $n$ variables have $n!$ possible orderings.}, not a typical context for machine learning classification.  Instead of the ordering itself, we aimed for machine learning to pick which of three existing heuristics \cite{DSS04}, \cite{BDEW13}, \cite{Brown2004} we should follow.  Experiments on over 7000 problems identified substantial subclasses on which each of the three heuristics made the best decision.  
This time we trained three SVM classifiers, one for each heuristic, and used the relative magnitude of their margin values to choose the one to follow for each problem.  We found this machine learned choice did significantly better than any one heuristic overall.

\section{Potential use in Symbolic Computation}
\label{SEC:OtherSC}

\subsection{Machine learning elsewhere in CAD/QE}
\label{SUBSEC:OtherQE}

It seems \cite{HEWDPB14} was the first publication on the application of machine learning to symbolic computation.  The only similar work since is \cite{KIMA16} which applied machine learning to decide the order of sub-formulae solving for their QE procedure\footnote{The feature set they used for their SVM was seeded from those in \cite{HEWDPB14}.}.  

There are certainly other decisions to be made when using CAD: such as the designation of equational constraints \cite{McCallum1999b}, \cite{EBD15}, \cite{BDEMW16}; and for some CAD algorithms even the order of polynomials and formulae \cite{EBCDMW14}.  Perhaps of most importance is the high level choice of which CAD implementation to use for a problem: most comparison experiments will show problem instances where different solvers prosper.  
Looking wider still, if the application problem were Quantifier Elimination then  there are a multitude of non-CAD approaches, such as virtual substitution \cite{Sturm2017} or QE by comprehensive GB \cite{FIS15b}, superior for classes of input.  
The author will be leading an upcoming EPSRC project (EP/R019622/1) on these topics.

\subsection{Machine learning elsewhere in computer algebra}
\label{SUBSEC:OtherCAS}

Computer Algebra Systems (CASs) often have a choice of algorithms to use when solving a problem.  Since a single one is rarely the best for the entire problem space, CASs usually use \emph{meta-algorithms} to choose, where decisions are often based on some numerical parameters \cite{Carette2004}.  A prominent example would be how and when to simplify mathematical expressions (see \cite{Stoutemyer2011} and references within).  Could machine learning be more effective?
In a presentation at ICMS 2016 it was reported that \textsc{Maple}'s user level symbolic integration command calls 16 different integration procedures in sequence until one returns an answer.  It is likely that the optimal order of calls would vary with problem instance.  Ever broader, a generic command like \textsc{Maple}'s \texttt{solve} or \textsc{Mathematica}'s \texttt{Solve} has to contend with not knowing exactly what the user means by ``\emph{solve}'', inferring from the input.  Machine learning could possibly assist with this, perhaps not just by viewing the input, but also the user's session history.

\section{Machine Learning elsewhere in Mathematical Software}
\label{SEC:MS}

\subsection{Satisfiability checking}
\label{SUBSEC:SAT}

There has been some use of machine learning within the satisfiability checking community for their SAT-solvers \cite{BHvMW09}.  These are tools dedicated to the solution of the Boolean SAT problem (given a Boolean formula decide if there is an allocation of values to variables that satisfies it).  Despite the SAT problem being NP-Complete, there exist solvers which can process formulae with millions of variables, and they are a common tool in many industries.

There is rarely a single dominant SAT solver; instead, different solvers perform best on different instances. The portfolio solver \textsc{SATZilla} \cite{XHHL08} takes sets of problem instances and solvers, and constructs a portfolio optimizing a given objective function (such as mean runtime, percent of instances solved, or score in a competition).  
\textsc{SATZilla} did well in SAT competitions\footnote{Although, because problems change little between competitions there is a risk of over-fitting being rewarded: \url{www.msoos.org/2018/01/predicting-clause-usefulness}}.

Machine learning within the actual search algorithms was a prominent part of the  \textsc{MapleSAT} \cite{LHPCG17} solver.  The developers view the question of solver branching as an optimisation problem where the objective is to maximize the learning rate, defined as the propensity for variables to generate learnt clauses. Experiments showed this to correlate well with efficiency, but the cost of an absolute solution could outweigh the savings.  Hence the chosen approach was to use machine learning to gain a heuristic solution to the optimisation problem.

Another use of machine learning in SAT is the choice of initial value to variable allocation to begin the search.  In \cite{Wu2017} the author describes using a logistic regression model to predict the satisfiability of  formulae after fixing the values of a certain fraction of the variables and adapting \textsc{MiniSAT} to determine the preferable initial values using this and a Monte-Carlo approach.  The author reported a high accuracy in the setting of backbone variables (variables that have the same value in all solutions of the formula) on initiation. 

\subsection{Satisfiability modulo theories}
\label{SUBSEC:SMT}

SAT-solvers can be applied to problems outside of the Boolean domain.  The approach, called Satisfiability Module Theories (SMT), is to iteratively use a SAT solver to find solutions to the Boolean skeleton of a formula and then query whether this is a solution in the  domain, learning new logical restrictions when not [Chapter 26]\cite{BHvMW09}.  In the domain of non-linear real arithmetic, symbolic algorithms developed for QE are the basis of these theory solvers and so the results and potentials in Sections \ref{SEC:MyWork} and \ref{SUBSEC:OtherQE} all apply.  There are likely similar questions of which tool to use for an instance in many of the other domains also. 

Machine learning can also be applied to fundamental questions regarding the Boolean encoding.  In \cite{SLB03} the authors studied whether it was best to encode atomic subformulas with Boolean variables, or to encode integer variables as bit-vectors, for working in separation logic with uninterpreted functions.  They concluded that a hybrid approach was needed after evaluating a wide range of benchmarks and used statistical techniques to decide what to do: an early application of a machine learning approach to SMT-solvers.

\subsection{Mathematical knowledge management}
\label{SUBSEC:MKM}

Perhaps the area of mathematical software with the greatest potential for machine learning applications is Mathematical Knowledge Management (MKM) \cite{CF09} since many of the tasks are similar to Natural Language Processing (NLP) where machine learning has seen extensive use.  For example, \cite{RS08} describes the automatic identification of a suitable top level from the  Mathematics Subject Classification (MSC) system for thousands of articles using an SVM; while \cite{SS14} describes how NLP techniques were adapted to build a part of speech tagger used for key phrase extraction in the database zbMATH.

\subsection{Automated reasoning}
\label{SUBSEC:AR}

Theorem Provers (TPs) prize correctness to a greater extent than even computer algebra systems.  They piece together mathematical results from the most basic rules of logic to give a certificate of correctness.  
The search space for proofs can be huge so we need techniques to cut it down or guide searches through.  So it is perhaps not surprising that Automated Reasoning has been looking at how best to use machine learning for some time.

The work surveyed in Section \ref{SEC:MyWork} followed \cite{BHP14} which used SVMs and Gaussian processes to select from different search strategies for the E prover (see references within for other studies).  Elsewhere, machine learning is used to select the most relevant theorems and definitions to use when proving a new conjecture in the \textsc{MaLARea} system \cite{Urban2007}.  An overview of such \emph{premise selection} approaches is given in \cite{KvTUH12} with the first deep learning approach detailed in \cite{ACEISU16}.

These approaches are relevant also for proof assistants.  For example, Sledgehammer allows for \textsc{Isabelle/HOL} to send goals to a variety of automated TPs and SMT solvers.  A relevance filter heuristically ranks the thousands of facts available and selects a subset based on syntactic similarity to the goal, with the \texttt{MaSh} option based on machine learning outperforming the standard \cite{KBKU13}.

\newpage

\section{Summary}

There are challenges in applying machine learning to mathematical software:
\begin{itemize}
\item Formulating choices in a way suitable for machine learning: e.g.~choosing from existing heuristics rather than an ordering directly (Section \ref{SUBSEC:VarOrd}).
\item Obtaining datasets of sufficient size for training: for the work in Section \ref{SUBSEC:GB} we had to build random polynomials while for that in Section \ref{SUBSEC:VarOrd} we borrowed benchmark sets from another discipline (SMT).
\item Making related choices in tandem: for example the best variable ordering for CAD may change after GB preconditioning! How best to deal with this?
\end{itemize}
However, we have described successful applications in diverse areas and noted some potentials $-$ an ICMS 2018 session should provide further inspiration. 

\subsubsection{Acknowledgements} 
Surveyed work in Section \ref{SEC:MyWork} was supported by 
the European Union's Horizon 2020 research and innovation programme under grant agreement No H2020-FETOPEN-2015-CSA 712689 (\textsf{SC}$^2$); and EPSRC grant EP/J003247/1.  The author is now supported by EPSRC grant EP/R019622/1.

\bibliographystyle{splncs03}
\bibliography{CAD}

\begin{thebibliography}{10}
\providecommand{\url}[1]{\texttt{#1}}
\providecommand{\urlprefix}{URL }

\bibitem{ACEISU16}
Alemi, A., Chollet, F., Een, N., Irving, G., Szegedy, C., Urban, J.: Deepmath -
  {D}eep sequence models for premise selection. In: Proc. 30th
  International Conference on Neural Information Processing Systems (NIPS '16). pp. 2243--2251. Curran Associates Inc. (2016)

\bibitem{ACM84I}
Arnon, D., Collins, G., McCallum, S.: Cylindrical algebraic decomposition {I}:
  The basic algorithm. SIAM J. Computing  13,  865--877 (1984)

\bibitem{BHvMW09}
Biere, A., Heule, M., {van Maaren}, H., Walsh, T.: Handbook of Satisfiability
  (Volume 185 Frontiers in Artificial Intelligence and Applications). IOS Press
  (2009)

\bibitem{BDEEGGHKRSW17}
Bradford, R., Davenport, J., England, M., Errami, H., Gerdt, V., Grigoriev, D.,
  Hoyt, C., Ko\v{s}ta, M., Radulescu, O., Sturm, T., Weber, A.: A case study on
  the parametric occurrence of multiple steady states. In: Proc. 2017 ACM International Symposium on Symbolic and Algebraic Computation (ISSAC '17). pp.
  45--52. ACM (2017)

\bibitem{BDEMW16}
Bradford, R., Davenport, J., England, M., McCallum, S., Wilson, D.: Truth table
  invariant cylindrical algebraic decomposition. J. Symbolic
  Computation  76,  1--35 (2016)

\bibitem{BDEW13}
Bradford, R., Davenport, J., England, M., Wilson, D.: Optimising problem
  formulations for cylindrical algebraic decomposition. In: Carette, J.,
  Aspinall, D., Lange, C., Sojka, P., Windsteiger, W. (eds.) Intelligent
  Computer Mathematics, LNCS, vol. 7961, pp.
  19--34. Springer Berlin Heidelberg (2013)

\bibitem{BHP14}
Bridge, J., Holden, S., Paulson, L.: Machine learning for first-order theorem
  proving. J. Automated Reasoning pp. 1--32 (2014)

\bibitem{Brown2004}
Brown, C.: Companion to the tutorial: {C}ylindrical algebraic decomposition,
  presented at {ISSAC} '04.
  \url{http://www.usna.edu/Users/cs/wcbrown/research/ISSAC04/handout.pdf}
  (2004)

\bibitem{BD07}
Brown, C., Davenport, J.: The complexity of quantifier elimination and
  cylindrical algebraic decomposition. In: Proc. 2007
  {I}nternational {S}ymposium on {S}ymbolic and {A}lgebraic {C}omputation (ISSAC '07). pp.
  54--60. ACM (2007)

\bibitem{BH91}
Buchberger, B., Hong, H.: Speeding up quantifier elimination by {G}r\"{o}bner
  bases. Tech. rep., 91-06. RISC, Johannes Kepler University (1991)

\bibitem{Carette2004}
Carette, J.: Understanding expression simplification. In: Proceedings of the
  2004 {I}nternational {S}ymposium on {S}ymbolic and {A}lgebraic {C}omputation (ISSAC '04).
  pp. 72--79. ACM (2004)

\bibitem{CF09}
Carette, J., Farmer, W.: A review of {M}athematical {K}nowledge {M}anagement.
  In: Carette, J., Dixon, L., Coen, C., Watt, S. (eds.) Intelligent Computer
  Mathematics, LNCS, vol. 5625, pp. 233--246.
  Springer-Verlag (2009)

\bibitem{CV95}
Cortes, C., Vapnik, V.: Support-vector networks. Machine Learning  20(3),
  273--297 (1995)

\bibitem{DBEW12}
Davenport, J., Bradford, R., England, M., Wilson, D.: Program verification in
  the presence of complex numbers, functions with branch cuts etc. In: 14th
  International Symposium on Symbolic and Numeric Algorithms for Scientific
  Computing. pp. 83--88. SYNASC '12, IEEE (2012)

\bibitem{DH88}
Davenport, J., Heintz, J.: Real quantifier elimination is doubly exponential.
  J. Symbolic Computation  5(1-2),  29--35 (1988)

\bibitem{DSS04}
Dolzmann, A., Seidl, A., Sturm, T.: Efficient projection orders for {CAD}. In:
  Proc. 2004 {I}nternational {S}ymposium on {S}ymbolic and
  {A}lgebraic {C}omputation (ISSAC '04). pp. 111--118. ACM (2004)

\bibitem{EBCDMW14}
England, M., Bradford, R., Chen, C., Davenport, J., {Moreno~Maza}, M., Wilson,
  D.: Problem formulation for truth-table invariant cylindrical algebraic
  decomposition by incremental triangular decomposition. In: Watt, S.,
  Davenport, J., Sexton, A., Sojka, P., Urban, J. (eds.) Intelligent Computer
  Mathematics, LNAI, vol. 8543, pp. 45--60.
  Springer International (2014)

\bibitem{EBD15}
England, M., Bradford, R., Davenport, J.: Improving the use of equational
  constraints in cylindrical algebraic decomposition. In: Proc. 2015 International Symposium on Symbolic and Algebraic Computation (ISSAC '15). pp.
  165--172. ACM (2015)

\bibitem{ED16a}
England, M., Davenport, J.: The complexity of cylindrical algebraic
  decomposition with respect to polynomial degree. In: Gerdt, V., Koepf, W.,
  Werner, W., Vorozhtsov, E. (eds.) Computer Algebra in Scientific Computing:
  18th International Workshop, CASC 2016, LNCS, vol. 9890, pp. 172--192. Springer (2016)

\bibitem{FIS15b}
Fukasaku, R., Iwane, H., Sato, Y.: Real quantifier elimination by computation
  of comprehensive {G}r\"{o}bner systems. In: Proc. 2015
  International Symposium on Symbolic and Algebraic Computation (ISSAC '15). pp. 173--180.  ACM (2015)

\bibitem{GS17}
Ghaffarian, S., Shahriari, H.: Software vulnerability analysis and discovery
  using machine-learning and data-mining techniques: {A} survey. ACM Comput.
  Surv.  50(4) (2017)

\bibitem{HEDP16}
Huang, Z., England, M., Davenport, J., Paulson, L.: Using machine learning to
  decide when to precondition cylindrical algebraic decomposition with
  {G}roebner bases. In: 18th International Symposium on Symbolic and Numeric
  Algorithms for Scientific Computing (SYNASC '16). pp. 45--52. IEEE (2016)

\bibitem{HEWDPB14}
Huang, Z., England, M., Wilson, D., Davenport, J., Paulson, L., Bridge, J.:
  Applying machine learning to the problem of choosing a heuristic to select
  the variable ordering for cylindrical algebraic decomposition. In: Watt, S.,
  Davenport, J., Sexton, A., Sojka, P., Urban, J. (eds.) Intelligent Computer
  Mathematics, LNAI, vol. 8543, pp.
  92--107. Springer International (2014)

\bibitem{KIMA16}
Kobayashi, M., Iwane, H., Matsuzaki, T., Anai, H.: Efficient subformula orders
  for real quantifier elimination of non-prenex formulas. In: Kotsireas, S.,
  Rump, M., Yap, K. (eds.) Mathematical Aspects of Computer and Information
  Sciences (MACIS '15). LNCS, vol. 9582, pp.
  236--251. Springer International (2016)

\bibitem{KBKU13}
K{\"u}hlwein, D., Blanchette, J., Kaliszyk, C., Urban, J.: {MaSh}: {M}achine
  learning for sledgehammer. In: Blazy, S., {Paulin-Mohring}, C., Pichardie, D.
  (eds.) Interactive Theorem Proving, LNCS, vol.
  7998, pp. 35--50. Springer Berlin Heidelberg (2013)

\bibitem{KvTUH12}
K{\"u}hlwein, D., {van Laarhoven}, T., Tsivtsivadze, E., Urban, J., Heskes, T.:
  Overview and evaluation of premise selection techniques for large theory
  mathematics. In: Gramlich, B., Miller, D., Sattler, U. (eds.) Automated
  Reasoning, LNCS, vol. 7364, pp. 378--392.
  Springer Berlin Heidelberg (2012)

\bibitem{LHPCG17}
Liang, J., {Hari Govind}, V., Poupart, P., Czarnecki, K., Ganesh, V.: An
  empirical study of branching heuristics through the lens of global learning
  rate. In: Gaspers, S., Walsh, T. (eds.) Theory and Applications of
  Satisfiability Testing -- {SAT} 2017, LNCS, vol.
  10491, pp. 119--135. Springer International (2017)

\bibitem{McCallum1999b}
McCallum, S.: On projection in {CAD}-based quantifier elimination with
  equational constraint. In: Proc. 1999 {I}nternational
  {S}ymposium on {S}ymbolic and {A}lgebraic {C}omputation (ISSAC
  '99). pp. 145--149. ACM (1999)

\bibitem{SS14}
Sch{\"o}neberg, U., Sperber, W.: {POS} tagging and its applications for
  mathematics. In: Watt, S., Davenport, J., Sexton, A., Sojka, P., Urban, J.
  (eds.) Intelligent Computer Mathematics, LNCS,
  vol. 8543, pp. 213--223. Springer International (2014)

\bibitem{SLB03}
Seshia, S., Lahiri, S., Bryant, R.: A hybrid {SAT}-based decision procedure for
  separation logic with uninterpreted functions. In: Proc. 2003
  Design Automation Conference. pp. 425--430 (2003)

\bibitem{SC04}
{Shawe-Taylor}, J., Cristianini, N.: Kernel methods for pattern analysis.
  CUP (2004)

\bibitem{Stoutemyer2011}
Stoutemyer, D.: Ten commandments for good default expression simplification.
  J. Symbolic Computation  46(7),  859--887 (2011)

\bibitem{Sturm2017}
Sturm, T.: A survey of some methods for real quantifier elimination, decision,
  and satisfiability and their applications. Math. in Comp. Sci.
  11(3),  483--502 (2017)

\bibitem{Urban2007}
Urban, J.: {MaLARea: A} metasystem for automated reasoning in large theories.
  In: Empirically Successful Automated Reasoning in Large Theories (ESARLT
  '07), CEUR Workshop Proceedings, vol. 257, 14 pages. CEUR-WS (2007)

\bibitem{RS08}
\v{R}eh\r{u}\v{r}ek, R., Sojka, P.: Automated classification and categorization
  of mathematical knowledge. In: Autexier, S., Campbell, J., Rubio, J., Sorge,
  V., Suzuki, M., Wiedijk, F. (eds.) Intelligent Computer Mathematics. LNCS, vol. 5144, pp. 543--557. Springer Berlin
  Heidelberg (2008)

\bibitem{WBD12_GB}
Wilson, D., Bradford, R., Davenport, J.: Speeding up cylindrical algebraic
  decomposition by {G}r\"{o}bner bases. In: Jeuring, J., Campbell, J., Carette,
  J., Reis, G., Sojka, P., Wenzel, M., Sorge, V. (eds.) Intelligent Computer
  Mathematics, LNCS, vol. 7362, pp. 280--294.
  Springer (2012)

\bibitem{Wu2017}
Wu, H.: Improving {SAT}-solving with machine learning. In: Proc.
  2017 ACM SIGCSE Tech. Symp. Computer Science Education. pp. 787--788. ACM (2017)

\bibitem{XHHL08}
Xu, L., Hutter, F., Hoos, H., {Leyton-Brown}, K.: {SAT}zilla: {P}ortfolio-based
  algorithm selection for {SAT}. J. Artificial Intelligence Research
  32,  565--606 (2008)

\bibitem{YSZ17}
Yadollahi, A., Shahraki, A., Zaiane, O.: Current state of text sentiment
  analysis from opinion to emotion mining. ACM Comput. Surv.  50(2) (2017)

\bibitem{YQKLK17}
Yau, K.L., Qadir, J., Khoo, H., Ling, M., Komisarczuk, P.: A survey on
  reinforcement learning models and algorithms for traffic signal control. ACM
  Comput. Surv.  50(3) (2017)

\bibitem{ZT03}
Zhang, D., Tsai, J.: Machine learning and software engineering. Software
  Quality J.  11(2),  pp. 87--119 (2003)

\end{thebibliography}

\end{document}